\documentclass[apj]{emulateapj}
\usepackage{graphicx,url,amssymb,amsmath,longtable,rotating,units,wasysym,listings,bm,color}
\usepackage{natbib,verbatim,enumerate}

\newcommand{\lp}{\left(}
\newcommand{\rp}{\right)}

\newcommand{\ee}{\end{eqnarray}}
\newcommand{\be}{\begin{eqnarray}}

\begin{document}
\normalsize

\slugcomment{Accepted for publication in The Astrophysical Journal}

\title{The Signature of Single-Degenerate Accretion Induced Collapse}

\author{Anthony L. Piro\altaffilmark{1} and Todd A. Thompson\altaffilmark{2}}

\altaffiltext{1}{Theoretical Astrophysics, California Institute of Technology, 1200 E California Blvd., M/C 350-17, Pasadena, CA 91125, USA; piro@caltech.edu}

\altaffiltext{2}{Department of Astronomy and Center for Cosmology \& Astro-Particle Physics, The Ohio State University, Columbus, OH 43210, USA}

\begin{abstract}
The accretion induced collapse (AIC) of a white dwarf to a neutron star has long been suggested as a natural theoretical outcome in stellar evolution, but there has never been a direct detection of such an event. This is not surprising since the small amount of radioactive nickel synthesized (\mbox{$\sim10^{-3}\,M_\odot$}) implies a relatively dim optical transient. Here we argue that a particularly strong signature of an AIC would occur for an oxygen-neon-magnesium (ONeMg) white dwarf accreting from a star that is experiencing Roche-lobe overflow as it becomes a red giant. In such cases, the $\sim10^{50}\,{\rm erg}$ explosion from the AIC collides with and shock-heats the surface of the extended companion, creating an \mbox{X-ray} flash lasting $\sim1\,{\rm hr}$ followed by an optical signature that peaks at an absolute magnitude of $\sim -16$ to $-18$ and lasts for a few days to a week. These events would be especially striking in old stellar environments where hydrogen-rich supernova-like, transients would not normally be expected. Although the rate of such events is not currently known, we describe observing strategies that could be utilized with high cadence surveys that should either detect these events or place strong constraints on their rates.
\end{abstract}


\section{Introduction}

As an accreting carbon-oxygen (CO) white dwarf (WD) grows toward the Chandrasekhar limit, a well-known potential outcome is ignition of its nuclear fuel, leading to a Type Ia supernova (SN Ia) \citep{Hillebrandt00}. However if the WD instead has a composition of oxygen-neon-magnesium (ONeMg), then the final outcome is strikingly different. In this case, electron captures on Ne and Mg rob the core of pressure support, causing the WD to collapse to a neutron star (NS) \citep{Canal76,Nomoto91}. This ``accretion induced collapse'' (AIC) has been invoked to explain millisecond pulsars \citep{Bhattacharya91}, subsets of gamma-ray bursts \citep{Dar92}, magnetars \citep{Usov92}, and proposed as a source of \mbox{$r$-process} nucleosynthesis \citep{Hartmann85,Fryer99}.

Despite its potential importance, there has been no reported detection of an AIC event. This is not too surprising since the expected AIC rate is no more than $\approx1\%$ of that of SNe Ia \citep{Yungelson98}. In addition, relative to Type I and Type II SNe, the ejecta mass is expected to be small ($\lesssim10^{-2}\,M_\odot$), high velocity ($\approx0.1c$), and produce little $^{56}$Ni ($\lesssim10^{-3}\,M_\odot$) \citep{Woosley92,Dessart06}. The resulting optical transient is thus considerably fainter than a typical SN (by 5 magnitudes or more) and lasts $\sim1\,{\rm d}$ (see our estimates later in this paper). If high angular momentum material forms a disk around the newly formed NS, this may increase both the ejecta mass and $^{56}$Ni yield by up to an order of magnitude \citep{Metzger09,Darbha10}. This requires a large amount of differential rotation in the WD just prior to collapse \citep{Abdikamalov10}, which is generally not expected \citep{Piro08}. If the AIC leads to a rapidly rotating magnetar, then other transient signatures in radio or X-rays might also occur \citep{Piro13,Metzger14}.

Many of the above scenarios focus on the merger of two CO WDs, with a combined mass greater than the Chandrasekhar mass, $M_{\rm Ch}\approx1.4\,M_\odot$, to produce an ONeMg WD, which then subsequently undergoes AIC. This is because the large angular momentum present in such systems may assist in generating a large magnetic field and powering an observable transient \citep[although see work by][which argues that the spin of the NS may not be that high after all]{Schwab12}. Nevertheless, accretion directly onto an ONeMg WD in a single-degenerate binary can also lead to AIC, and these WDs are present as a natural by-product of the evolution of stars with masses in the (uncertain) range of $\sim6-8\,M_\odot$ \citep[or even up to $\sim10\,M_\odot$;][and references therein]{Garcia97}. There is in fact direct evidence for ONeMg WDs from the composition of nova ejecta \citep{Truran86,GilPons03} and the high masses ($\gtrsim1.1\,M_\odot$) of a subset of field WDs \citep{Baxter14}. These high-mass WDs have been shown to lead to AIC in single-degenerate systems with a wide range of donors, including main-sequence stars, red giants, and helium stars \citep[][and references therein]{Tauris13}, and may even lead to millisecond pulsars in eccentric orbits \citep{Freire14}.

For the present study, we focus on the particular case of an ONeMg WD accreting from a $\approx0.9\,M_\odot$ companion that is undergoing Roche-lobe overflow as it ascends the red-giant branch. Such a scenario is expected in old field populations where, without interactions from a dense stellar environment, this will be the primary way for initiating mass transfer. Furthermore, the expected accretion rate is in the correct range to lead to AIC \citep{Tauris13}. As we show below, this large companion is especially useful for generating a bright transient from interaction with the AIC explosion. These transients should have properties that are unexpected in old stellar populations, which should help in identifying them uniquely in surveys.

In Section \ref{sec:masstransfer}, we describe the expected mass transfer scenarios, which will set the range of separations expected for the binary at the moment of AIC. In Section \ref{sec:observables}, we provide estimates of the range of luminosities expected for the AIC signature along with calculations of example light curves. In Section \ref{sec:rates} we summarize the expected rate for these events. We conclude in Section \ref{sec:conclusion} with a summary of our work and a discussion of potential strategies for detection.

\section{Binary Evolution}\label{sec:masstransfer}

In this section, we consider the time-evolution of an ONeMg WD accreting from a red giant star. This sets the typical accretion rate in such systems and also the separation at the moment of AIC. These factors are important for determining the luminosity of the associated optical transient in Section \ref{sec:observables}. We focus on old field stellar populations, which significantly limits the range of systems we must investigate because only companions with masses of $M_2\approx0.9\,M_\odot$ will be presently moving away from the main sequence. We also only have to consider a Population I metallicity, consistent with studies of elliptical galaxies \citep{Trager00}.

To become an ONeMg WD, the primary of the binary system begins with a zero age main sequence mass of \mbox{$\sim6.5-8\,M_\odot$}. When the primary leaves the main sequence, our scenario requires that a common envelope phase is initiated. This is needed to shrink the binary so that the companion can later overflow its Roche-lobe as it becomes a red giant. The are two potential opportunities for initiating common envelope, (1) when the primary ascends the red giant branch (RGB), and (2) when the primary ascends the asymptotic giant branch (AGB).

We explore the maximum radius expected in each case by running stellar models with the stellar evolution code \texttt{MESA} \citep{Paxton11} and using the binary evolution code \texttt{BSE} \citep{Hurley02}. When using \texttt{MESA}, we adopted the Reimers mass loss formula with $\eta_R=0.5$ for the RGB phase \citep{Reimers75}, while for the AGB phase we used the mass loss formula of \citet{Bloecker95} with a range of $\eta_B=0.05-0.5$.  We find that the primary inflates to $\approx200-300\,R_\odot$ during the RGB with both \texttt{MESA} and \texttt{BSE}. The AGB is a little more difficult, and in particular, \texttt{MESA} can potentially have issues with thermal pulses depending on the exact mass loss value used. Nevertheless, over the parameter range considered, \texttt{MESA} gives maximum radii during the AGB just shy of $\approx100\,R_\odot$ and  \texttt{BSE} gives $\approx1400\,R_\odot$. Thus this robustly shows that during the AGB the star will be a factor of $\sim4$ larger than during the RGB, and there is suitable parameter space where the binary will not experience common envelope during the RGB, but will during the AGB.

There are a couple of constraints on the binary evolution to correctly produce the correct final system. First, the binary must avoid common envelope during the RGB to make sure an ONeMg WD is produced. Given the RGB maximum radii of $\approx200-300\,R_\odot$, the initial orbital period must then be greater than $\approx350-700\,{\rm d}$ to survive the RGB.  Next, to make sure that Roche-lobe overflow occurs during the AGB phase, the initial orbital period must be less than $\approx7000\,{\rm d}$. Using typical common envelope prescriptions \citep{deKool90}, we then estimate the new orbital period once the primary's envelope is ejected as $\approx10-500\,{\rm d}$. This large range is mainly due to differences in mass loss during the AGB phase and the treatment of the common envelope phase. This new period is sufficiently short that the companion will experience Roche-lobe over flow when it leaves the main sequence. Using this as a starting point, in the following we consider the orbital period $P_0$ at the moment of Roche-lobe contact to be a free parameter and investigate the results for a range of values.

\begin{figure}
\epsscale{1.2}
\plotone{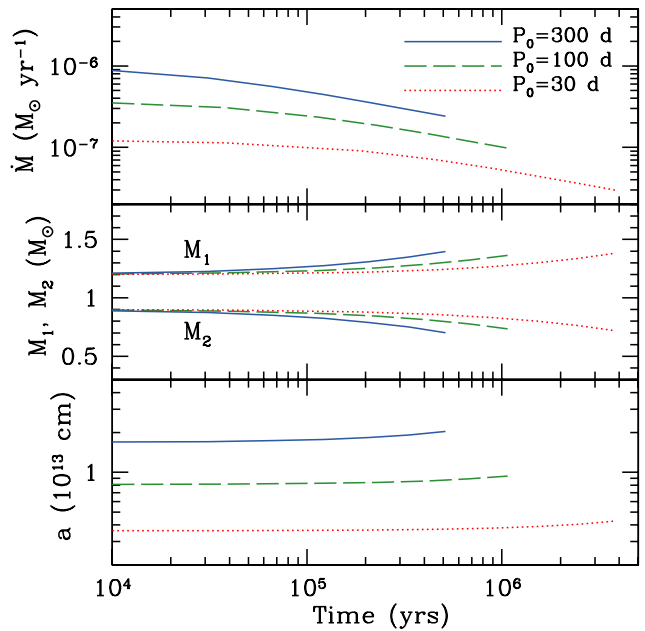}
\caption{Time-dependent binary evolution of the accretion rate $\dot{M}$, the masses of the primary and companion $M_1$ and $M_2$, respectively, and the orbital separation $a$. Roche-lobe contact occurs at $P_0$ (for which we compare $30\,{\rm d}$, $100\,{\rm d}$, and $300\,{\rm d}$, as denoted) and proceeds until \mbox{$M_1=M_{\rm Ch}$}, at which point AIC occurs.}
\label{fig:aic_evolution}
\epsscale{1.0}
\end{figure}

The subsequent mass transfer is calculated using the framework described in \citet{Ritter99}, where the companion expands as it becomes a red giant and always remains in thermal equilibrium since the mass transfer is relatively slow. Since the radius of the companion $R_2$ must remain in contact with its Roche lobe,
\be
	R_2=R_{\rm R,2}\approx 0.46\lp \frac{M_2}{M_1+M_2}\rp^{1/3}a,
\ee
where $a$ is the orbital separation, all parameters of the binary can be expressed in term of the initial orbital period \citep{Piro02}. In this way, we solve for the mass transfer rate
\be\label{eq:mdot}
	\dot{M} = 2.7\times10^{-8}\lp\frac{P_0}{100\,{\rm d}}\rp^{14/15}
	\lp\frac{M_1}{1.2\,M_\odot}\rp^{-14/15}
	\nonumber
	\\
	\times
	\lp\frac{M_2}{0.9\,M_\odot}\rp^{-7/15}
	\lp \frac{5}{6}\frac{M_\odot}{M_2}-\frac{M_\odot}{M_1}\rp^{-1}
	M_\odot\,{\rm yr}^{-1}.
\ee
This naturally gives accretion in the range of \mbox{$\gtrsim10^{-7}\,M_\odot\,{\rm yr}^{-1}$}.

For understanding the fate of the accreting WD, there are three key accretion rates to consider. If the accretion rate exceeds $\dot{M}_{\rm steady}\approx3\times10^{-7}\,M_\odot\,{\rm yr^{-1}}$ (where the exact value is set by the relatively high mass of the accreting WD), the accreted hydrogen-rich fuel can steadily burn to helium \citep[][and references therein]{Wolf13}. If the accretion rate is below $\dot{M}_{\rm steady}$, but still above $\dot{M}_{\rm weak}\approx 10^{-7}\,M_\odot\,{\rm yr^{-1}}$, there are likely recurrent shell flashes, but these are too weak to prevent overall mass accumulation \citep{Hachisu99}. Finally, if the accretion rate gets above $\dot{M}_{\rm giant}\approx10^{-6}\,M_\odot\,{\rm yr^{-1}}$, then  the accreting WD may puff up to become a giant with associated winds preventing mass accumulation \citep{Nomoto07,Shen07}.

Using the accretion rate of Equation (\ref{eq:mdot}), and assuming mass and angular momentum conservation, we integrate the binary forward in time until $M_1$ exceeds the Chandrasekhar mass $M_{\rm Ch}$ at which point we assume that electron captures are initiated and the core collapses. In Figure \ref{fig:aic_evolution}, we plot example binary evolution calculations of such systems. To be able to stably accumulate fuel and reach $M_{\rm Ch}$ requires $\dot{M}_{\rm weak}<\dot{M}<\dot{M}_{\rm giant}$.  We find that this is only satisfied for a relatively narrow range of initial periods of $P_0\approx100-300\,{\rm d}$ \citep[see also][]{Tauris13}, which implies a relatively small diversity in the possible optical signatures.

The final period, or separation, is simply set by angular momentum conservation
\be
	P = \lp\frac{M_{1,0}}{M_1}\rp^3\lp\frac{M_{2,0}}{M_2}\rp^3P_0, 
\ee
where $M_{1,0}$ and $M_{2.,0}$ are the initial masses of the primary and companion, respectively. For typical parameters, we therefore find that $P$ increases by $\approx30\%$ from $P_0$ and thus $a$ may increase by $\approx20\%$. For a maximally spinning WD, assuming solid-body rotation, the Chandrasekhar mass may be increased to $1.48\,M_\odot$ \citep{Yoon05}. In this case $a$ may increase by $\approx40\%$. This demonstrates that the initial period is roughly setting the separation at the moment of AIC.

\section{Shockwave Collision with the Red Giant Companion}\label{sec:observables}

\begin{figure}
\epsscale{1.2}
\plotone{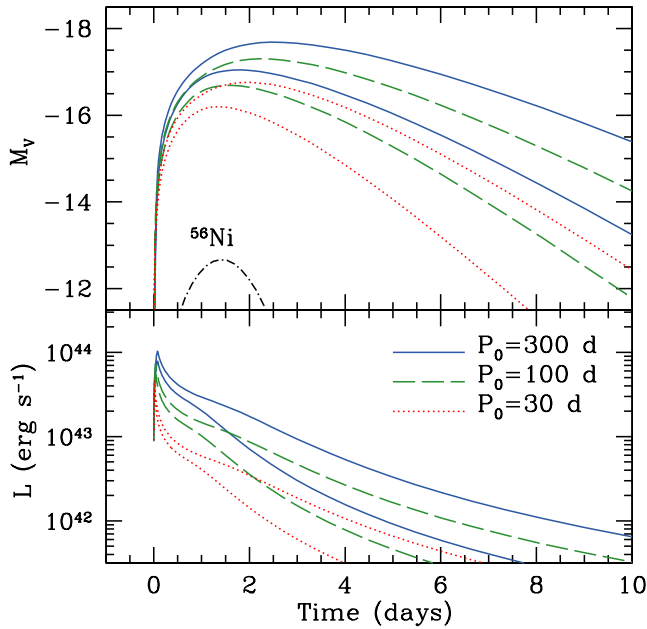}
\caption{The upper panel plots $V$-band absolute magnitude versus time after the collision for cooling of the shock-heated companion (three different initial periods are compared) and $^{56}$Ni emission (labeled and using a dot-dashed line). The bottom panel is the bolometric luminosity from the cooling of the shock heated companion. In eafch case, the lower (upper) curve corresponds to $M_{\rm tot}$ of $10^{-3}\,M_\odot$ ($3\times10^{-3}\,M_\odot$).}
\label{fig:lightcurve}
\epsscale{1.0}
\end{figure}

Although an AIC results in the majority of the WD imploding and forming an NS, core-bounce leads to an outgoing shock that is powered by neutrinos to create a successful, albeit weak, supernova-like explosion \citep{Woosley92,Dessart06}. Typical parameters are energies of $\sim10^{50}\,{\rm erg}$, with an ejecta mass of $M_e\sim3\times10^{-3}\,M_\odot$ and synthesizing $\sim10^{-3}\,M_\odot$ of $^{56}$Ni (although we note that the models of \citealp{Fryer99} and \citealp{Fryer09} produce significantly more ejecta and $^{56}$Ni). In the upper panel of Figure \ref{fig:lightcurve}, we plot an estimated $V$-band light curve of such an event (dot-dashed line), using the simple model presented in \cite{Li98} and including an additional exponential suppression of the luminosity when the ejecta becomes optically thin (further discussed below). The temperature is assumed to be a black body at each time. Even with these simple models it is clear that it is difficult to generate a bright transient from just this emission alone.

A key feature of this single-degenerate AIC model is the large companion that is nearby at the moment of AIC. The AIC explosion collides with this large target, heating it and producing an additional transient signal. Using the analytic results of \citet{Wheeler75}, we estimate that $\sim10^{-4}-10^{-2}\,M_\odot$ may be ejected from the companion when this happens, with ablation dominating over mass stripping \citep[also see][]{Pan12}. The light curve from such a process was investigated by \citet{Kasen10} for the case of an SN Ia colliding with its companion. The first possible emission will be X-rays if photons from the shock interaction can escape though the hole carved out by the collision,
\be\label{eq:lum}
	L_x = 3\times10^{44}\lp \frac{M_e}{10^{-3}\,M_\odot}\rp^{1/2}
		\lp\frac{v_e}{0.1c}\rp^{5/2}{\rm erg\,s^{-1}},
\ee
where $v_e$ is the ejecta velocity. This lasts roughly the shock crossing time of $R_2/v_e\approx1\,{\rm hr}$. Since the luminosity of this emission is independent of the companion radius (in contrast to the optical signature described next), although it is an important confirmation of the picture described here, it is not a particularly useful diagnostic for constraining the properties of the companion.

Following the X-ray flash, a longer-lasting optical transient is expected from cooling of deeper, shock heated material. Scaling the analytic estimates of \citet{Kasen10} to the typical values for an AIC results in a luminosity
\be\label{eq:lum}
	L = 10^{43}\lp\frac{a}{10^{13}\,{\rm cm}}\rp \lp \frac{M_e}{10^{-3}\,M_\odot}\rp^{1/4}
		\lp\frac{v_e}{0.1c}\rp^{7/4}
	\nonumber\\
	\times
	\lp \frac{t}{1\,{\rm d}}\rp^{-1/2}{\rm erg\,s^{-1}},
\ee
with an effective temperature of
\be\label{eq:teff}
	T_{\rm eff} = 2.5\times10^4 \lp\frac{a}{10^{13}\,{\rm cm}}\rp^{1/4}
	\lp \frac{t}{1\,{\rm d}}\rp^{-37/72}{\rm K}.
\ee
One detail these scalings do not account for is the possible recombination of the ejecta as it expands and cools \citep{Kleiser14}. For the relatively hot temperature given by Equation (\ref{eq:teff}), this occurs well after peak and does not impact our peak magnitude estimates.

Unlike the SN Ia case, this cooling of the shock heated, expanding material is never overtaken by the $^{56}$Ni produced in the main SN event. Therefore we have to take care for when this material is expected to become optically thin. The optical depth of the material heated and excavated from the red giant is roughly $\tau \approx 3M_{\rm tot}\kappa/4\pi (v_et)^2$, where $\kappa$ is the opacity and $M_{\rm tot}$ is the total mass ejected from the AIC and ablation of the companion. This shows that the material becomes optically thin on a timescale of
\be
	t_{\tau=1} \approx 2 \lp \frac{M_{\rm tot}}{10^{-3}\,M_\odot}\rp^{1/2} \lp\frac{v_e}{0.1c}\rp^{-1}{\rm d},
	\label{eq:ttau1}
\ee
where we estimate $\kappa=0.34\,{\rm cm^2\,g^{-1}}$, as is appropriate for electron scattering in solar composition material. It is likely that the ejecta expands from the red giant with velocities $\lesssim v_e$, and it may be even $\ll v_e$ if the matter is dominated by ablated material. Therefore, Equation~(\ref{eq:ttau1}) gives a conservative lower limit to the time when the material becomes optically thin. When $\tau\lesssim1$, then the internal energy of the ejecta streams out on roughly a light crossing time $t_{\rm lc}\approx r/c$. Since the luminosity in Equation (\ref{eq:lum}) was derived in the optically thick limit, the optically thin limit can be estimated by multiplying it by the ratio $t_{\rm diff}/t_{\rm lc}$ where $t_{\rm diff}\approx 3M_e\kappa/4\pi cv_et$ is the diffusion time. A quick comparison shows that this ratio is simply $\tau$, thus we include a factor of $1-e^{-\tau}$ for the luminosity, since this correctly goes from a value of $1$ when $\tau\gg1$ to $\approx\tau$ when $\tau\ll1$.

In Figure \ref{fig:lightcurve}, we plot example light curves using Equations (\ref{eq:lum}) and (\ref{eq:teff}), where we have assumed that the emission is roughly black body. Since there is uncertainty in exactly how much mass is ejected, we compare two different $M_{\rm tot}$ for each $P_0$, providing an idea of the expected luminosities and timescales. For $P_0\approx100-300\,{\rm d}$, where we expect mass accumulation and AIC, we find a rather narrow range of peak absolute magnitudes around $\sim-17$. We also include the case of $P_0=30\,{\rm d}$ to show a dimmer example, although we believe this is disfavored since, at least in standard accounts of WD accretion, such a system will undergo shell flashed and is not expected to accumulate sufficient mass to produce an AIC.

The strength of the optical and X-ray emission will depend on the viewing angle, with the strongest signal coming when the observer is looking down upon the shocked region. SNe Ia studies find the shock emission is viewable $\sim10\%$ of the time, but this will be higher for AIC since there is significantly less ejecta. Hopefully our work motivates numerical studies using multi-dimensional radiation-hydrodynamics calculations to address the viewing angle corrections in detail.

\section{Event Rates}\label{sec:rates}

The rate of such events is not currently known because there has been no smoking gun detection of an AIC. Nevertheless, rates and rate constraints have been discussed many times before in the literature \citep[][and references therein]{Livio01,Ruiter10}. Since AIC can result from similar channels to those popularly discussed in the literature for SNe Ia (namely single- and double-degenerate), it makes sense to think of the AIC rate in relation to the SN Ia rate. The Lick Observatory Supernova Search finds a rate of $(3.01\pm0.062)\times10^{-5}\,{\rm Ia\,Mpc^{-3}\,yr^{-1}}$ \citep{Li11}, which corresponds to $(4.0-7.1)\times10^{-3}\,{\rm Ia\,yr^{-1}}$ for the Milky Way. Using population synthesis, \citep{Yungelson98} find AIC rates of $8\times10^{-7}-8\times10^{-5}\,{\rm yr^{-1}}$ for the Milky Way, depending on assumptions about the common-envelope phase and mass transfer. Alternative constraints on AIC have been made from nucleosynthetic yields, and in particular, on the neutron-rich isotopes expected from these events \citep{Hartmann85,Woosley92,Fryer99}. These give upper limits similar to the population synthesis rate predictions.

For the specific scenario presented here, we are interested in just a subset of all AICs, since double degenerate scenarios will not produce the transient signature we predict. Thus it is a useful exercise to at least provide a very rough estimate of the expected rate for these events. For a Salpeter initial mass function \citep{Salpeter55}, $\sim1\%$ of stars have a zero age main sequence mass from $6$ to $8\,M_\odot$. Assuming a flat probability distribution of companion masses \citep{Duchene13} and a log normal distribution of orbital separations \citep{Abt83}, $\sim10\%$ have companions of mass around $\approx0.9\,M_\odot$ and about $\sim10\%$ have initial orbital periods in the needed range of $\approx300\,{\rm d}$ to $\approx7000\,{\rm d}$ (from the discussion in Section \ref{sec:masstransfer}), respectively. Combining this with $\sim50\%$ binaries \citep{Lada06,Kobulnicky07}, we estimate a rate of $\sim5\times10^{-5}\,{\rm yr^{-1}}$ for a Milky Way-like galaxy, similar to the constraints summarized above.

\section{Conclusion and Prospects for Detection}\label{sec:conclusion}

We have investigated the observational signature of an AIC occurring in a single degenerate system where a ONeMg accretes from a $0.9\,M_\odot$ star that is ascending the red giant branch. We show that when the weak supernova of the AIC collides with the companion, we expect an X-ray flash with lasting $\sim1\,{\rm hr}$ and a bright optical transient with a peak absolute magnitude of $\sim-16$ to $-18$ and lasting for a few days to a week. We argue that a short timescale hydrogen-rich, supernova-like transient observed in an old stellar population may be an especially strong signature of AIC. For SNe Ia, constraints on mass stripping have been made down to a level of $\sim10^{-3}\,M_\odot$ of hydrogen \citep{Shappee13a}, demonstrating that the amount of hydrogen we expect stripped in AIC events will be detectable.

In principle, these events could have been detected by wide-field, transient surveys like the Palomar Transient Factory \citep[PTF;][]{Rau09} and the Panoramic Survey Telescope and Rapid Response System \citep[Pan-STARRS;][]{Kaiser02}, and their apparent non-detections (assuming that these have not been detected but discarded because they are not bring searched for or are only seen in one epoch) should place upper limits on the rate. Unfortunately, there are not many precedents for transients with similar luminosities and timescales, so it is not clear how robust such constraints are. This should change in the future if there are surveys with particularly rapid cadences of $\sim1\,{\rm day}$ such as the Zwicky Transient Facility \citep[ZTF;][]{Law09}, the All-Sky Automated Survey for Supernovae \citep[ASAS-SN;][]{Shappee13b}, or the Large Synoptic Survey Telescope \citep{LSST}.

For example, consider a scenario where ZTF could cover $3800\,{\rm deg^2\,hr^{-1}}$ down to $m=20.4$ with 30 second exposures (E. Bellm, private communication). Over a 4~hour period $\sim37\%$ of the sky would be covered, which could then be repeated so that over an 8 hour night there would be 2 data points for each location. Assuming a threshold for detection at $M=-17$, this would allow events to be detected out to $\sim300\,{\rm Mpc}$. For a rate of $\sim3\times10^{-7}\,{\rm yr^{-1}\,Mpc^{-3}}$ (roughly $\sim1\%$ of the SNe Ia rate as described in Section \ref{sec:rates}), this gives $\sim35$ events per year within the observable volume. Finally, multiplying by the $37\%$ sky coverage per night we estimate $\sim13$ events could be detected per year. Even though this estimate will be affected by what fraction of AICs actually occur via a single-degenerate channel and how strong the viewing angle effects are (see our discussion at the end of Section \ref{sec:observables}), it demonstrates that interesting constraints on the rate of these events, or, possibly, a discovery, should be possible in the coming years.

\acknowledgements
We thank Drew Clausen for assistance with \texttt{MESA} to generate RGB and AGB models and discussions about binary evolution scenarios. We thank Selma de Mink for insight on binary interactions, Eric Bellm for feedback on rates, and Josiah Schwab for help with \texttt{MESA}. We also thank Edo Berger, Ryan Chornock, Dan Kasen, Christopher Kochanek, Brian Metzger, and Christian Ott for comments on previous drafts. We thank the Center for Cosmology and Astro-Particle Physics for funding ALP's visit to Ohio State University, where this work began. ALP is supported through NSF grants AST-1205732, PHY-1068881, PHY-1151197, and the Sherman Fairchild Foundation.

\bibliographystyle{apj}
\bibliography{ms}
\end{document}